\newcommand{\bea}{\begin{eqnarray}}
\newcommand{\eea}{\end{eqnarray}}
\newcommand{\nn}{\nonumber}

\documentclass{ws-ijmpa}
\usepackage[super,compress]{cite}
\usepackage{graphicx}
\begin{document}
\markboth{A. Kuiroukidis}{Inflationary $\alpha-$attractors and $F(R)$ gravity }

%
\catchline{}{}{}{}{}
%

\title{Inflationary $\alpha-$attractors and $F(R)$ gravity
}

\author{A. Kuiroukidis
}

\address{Department of Informatics,\\
Technological Education Institute of Central Macedonia,\\
GR 621 24, Serres Greece\\
apostk@teicm.gr}



\maketitle

\begin{history}
\received{Day Month Year}
\revised{Day Month Year}
\end{history}

\begin{abstract}
We consider a generic class of the so-called
inflationary $-\alpha$ attractor models and compute the
cosmological observables in the Einstein and Jordan frames,
of the corresponding $F(R)-$gravity theory. We find that
the two sets coincide (to within errors from the use of the
slow-roll approximation) for moderate and large values of
the number of e-foldings $N$, which is the novel result
of this paper, generalizing previous results on the subject
(see e.g. Ref. (\citen{oik1})). We briefly comment on the
possible generalizations of these results.

\end{abstract}

\ccode{PACS numbers:}





\section{Introduction}

Inflationary cosmology is the main theoretical description of the
Early Universe in the context of which it was possible to address
and solve the main theoretical problems of the Standard Big Bang
description of our Universe (see Refs. [\citen{mukh}]-[\citen{lyth}]).
Other theoretical attempts to solve the Early Universe puzzles
include the so-called bouncing cosmological models
(see Refs. [\citen{nove}]-[\citen{oik2}]).
In what follows in this paper we will be comparing our results
for the cosmological observables, namely the spectral index of primordial
curvature perturbations $n_{s}$ and the tensor-to-scalar ratio $r$,
with those of the Planck observational data \cite{ade}.

Recently an interesting class of models was discovered
in Ref. [\citen{lind1}], called the $\alpha-$attractors models with the
property that the cosmological observables are identical for
all the members of the $\alpha-$class, in the large $N-$limit,
where $N$ is the number of e-foldings. Subsequently these models
were studied more extensively (see Refs. [\citen{ferr}]-[\citen{zyi}]).
Also a recent study is that of Ref. [\citen{oik1}] which the present
study follows and is a generalization of it.

The above $\alpha-$attractor inflationary potentials have a large
flat plateau for large values of the inflaton scalar field and in the
small $\alpha-$limit are asymptotically quite similar to the hybrid
inflation scenarios \cite{kall3}. Well known inflationary models
are special limiting cases of the $\alpha-$attractors models,
such as the Starobinsky model \cite{star}.

In this paper we compute the cosmological observation parameters
of the spectral index of primordial curvature perturbations
$n_{s}$ and the tensor-to-scalar ratio $r$ for a more general class
of inflationary potentials with the $\alpha-$attractors property,
than those examined in Refs. [\citen{oik1}],[\citen{lind1}].
This potential was suggested in \cite{kall1}. More specifically
we compute these cosmological observables in the so-called Einstein
frame \cite{oik1}, where the form of the potential in explicitly
given in the action functional. Then we find the same cosmological
observables \cite{hnoh} in the so-called Jordan frame, where the
action functional is that of the corresponding $F(R)-$gravity theory
(see Refs. [\citen{capo}]-[\citen{seba}])
and we explicitly compute the corresponding $F(R)$ gravity theory.

Then we want to compare these two sets of values for the cosmological
observables in order to test \cite{oik1} whether the two frames' descriptions
are equivalent observationaly. The equivalence of the descriptions in the two
frames was explicitly shown in Ref. (\citen{derue}). See however
Ref. (\citen{capo2}) for an important exception. We thus find that these
two sets of observables coincide (to within
computational errors from the use of the so-called slow-roll approximation),
for moderate and large values of the number of e-foldings $N$,
a result that generalizes those of Ref. [\citen{oik1}], in a novel way.
This occurs for reasonable values of the set of parameters of the
inflationary potential and for small to moderate $\alpha-$values.
For more references related to inflation in $F(R)$ gravity
theories see Refs. [\citen{noji1}]-[\citen{maka}].

This paper in organized as follows: In section 2 we refer to the basic
facts about inflationary $-\alpha$ attractor models and compute the
corresponding $F(R)-$gravity theory. In section 3 we compute the
cosmological observables in the Einstein frame and in section 4
in the Jordan frame. Finally in section 5 we present the results
and discuss them.

In this paper we assume that the metric of the cosmological spacetime
is that of the flat FRW model
\bea
\label{frw}
ds^{2}=-dt^{2}+a^{2}(t)[dx^{2}+dy^{2}+dz^{2}]
\eea
where $a(t)$ is the scale factor and $\dot{H}=\dot{a}/a$ is the
Hubble parameter. Also we assume that the connection is a
symmetric, metric compatible and torsion-less affine connection,
namely the so-called Leci-Civita connection and the corresponding
Ricci-scalar curvature is given by
\bea
\label{ricci}
R=6(\dot{H}+2H^{2})
\eea
Finally we use units where $\kappa^{2}:=8\pi G=1$ and $h=c=1$.

\section{Basics of Inflationary $\alpha-$attractors and $F(R)$ gravity}

In this section we introduce the basic discussion concerning the
inflationary $\alpha-$attractor models, which are classes of
inflationary potentials with large flat potential plateaus, and their
relation to the cosmological observables. Thus we consider
the $F(R)$ gravity action in the so-called Jordan frame
\bea
\label{act1}
\hat{S}=\frac{1}{2}\int d^{4}x\sqrt{-\hat{g}}F(\hat{R})
\eea
where $\hat{g}_{\mu\nu}$ is the metric in the Jordan frame and $\hat{R}$
the corresponding Ricci scalar curvature. Introducing the auxiliary
scalar field $A$, one can write the above action as\cite{oik1}
\bea
\label{act2}
\hat{S}_{1}=\frac{1}{2}\int d^{4}x\sqrt{-\hat{g}}[F^{'}(A)(\hat{R}-A)+F(A)]
\eea
By varying the action $\hat{S}_{1}$ with respect to $A$ we obtain
$A=\hat{R}$ and verify thus the equivalence of the actions
(\ref{act1}), (\ref{act2}). Making the conformal transformation
\bea
\label{conf}
\hat{g}_{\mu\nu}=e^{\phi}g_{\mu\nu}
\eea
and introducing the canonical transformation
\bea
\label{cano}
\Phi=\sqrt{3}\phi\nn \\
\frac{e^{\phi}F^{'}(A)}{2}=1
\eea
corresponding to Eq. (20) of Ref. \citen{oik1}, the action is transformed
to $\hat{S}_{1}\longrightarrow S_{1}$, namely to the action in the Einstein
frame
\bea
\label{act3}
S_{1}=\int d^{4}x\sqrt{-g}[R-\frac{1}{2}g^{\mu\nu}\partial_{\mu}\Phi\partial_{\nu}\Phi-V(\Phi)]
\eea
where
\bea
\label{pot1}
V(\Phi):=2\left[\frac{A}{F^{'}(A)}-\frac{F(A)}{(F^{'}(A))^{2}}\right]=\nn
\\
=e^{\Phi/\sqrt{3}}\hat{R}(e^{-\Phi/\sqrt{3}})-\frac{1}{2}e^{2\Phi/\sqrt{3}}
F[\hat{R}(e^{-\Phi/\sqrt{3}})]
\eea
Here the dependence of $\hat{R}$ on $\Phi$ is found by solving the second
of Eqs. (\ref{cano}) with respect to $A=\hat{R}$. Also we obtain from this
\bea
\label{deri}
\frac{d\Phi}{d\hat{R}}=-\sqrt{3}\frac{F^{''}(\hat{R})}{F^{'}(\hat{R})}
\eea
and using Eq. (\ref{pot1})we finally obtain (see Eq. (24) of Ref. (\citen{oik1}))
\bea
\label{rel1}
\hat{R}F_{\hat{R}}=-2\sqrt{3}\frac{d}{d\Phi}
\left[\frac{V(\Phi)}{e^{2\Phi/\sqrt{3}}}\right]
\eea
In our conventions and notation we have $-\infty<\Phi\leq 0$ for the
scalar field. For example, for the Starobinsky model \cite{star}, which
in our notational conventions is given by
\bea
\label{sta1}
V(\Phi)=\mu^{2}(1-e^{\Phi/\sqrt{3}})^{2}
\eea
we obtain the corresponding $F(\hat{R})$ gravity description as \cite{oik1}
\bea
\label{frg1}
F(\hat{R})=\hat{R}+\frac{\hat{R}^{2}}{4\mu^{2}}
\eea

The above action of Eq. (\ref{act3}) can also occur from an action,
in the so-called $\phi-$Jordan frame\cite{oik1}, with
a non-canonically coupled scalar field of the form
\bea
\label{act4}
S_{n}=\int d^{4}x\sqrt{-g}
\left[R-\frac{\partial_{\mu}\phi\partial^{\mu}\phi}{2\left(1-\frac{\phi}{6\alpha}\right)^{2}}-V(\phi)\right]
\eea
Making the transformation
\bea
\label{transf}
\frac{d\phi}{\left(1-\frac{\phi}{6\alpha}\right)}=d\Phi
\eea
we finally obtain the action of Eq. (\ref{act3}), namely
\bea
\label{act5}
S_{1}=\int d^{4}x\sqrt{-g}
\left[R-\frac{1}{2}(\partial \Phi)^{2}-V\left(\sqrt{6\alpha}tanh\left(\frac{\Phi}{\sqrt{6\alpha}}\right)\right)\right]
\eea

\begin{figure}[h!]
\centerline{\includegraphics[width=10.8cm]{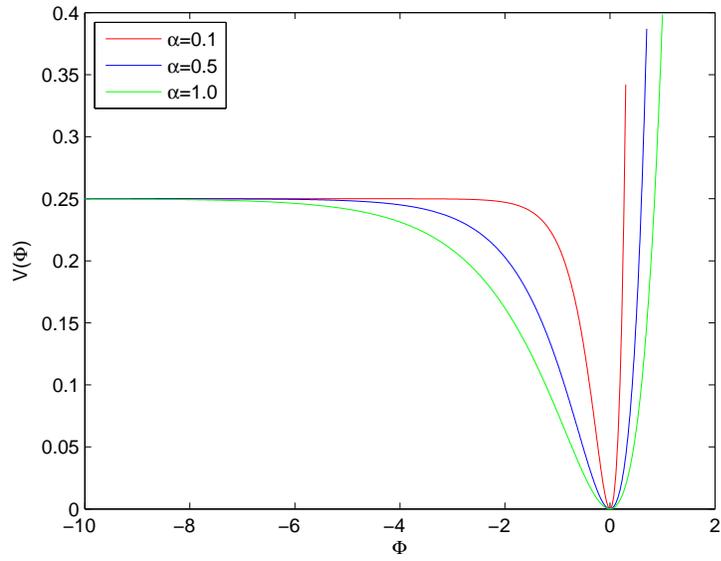}}
\caption{The potential of Eq. (\ref{pot2}) for $V_{0}=1$, $n=m=1$,
for various values of the parameter $\alpha$.
\label{fig1}}
\end{figure}

\begin{figure}[h!]
\centerline{\includegraphics[width=10.8cm]{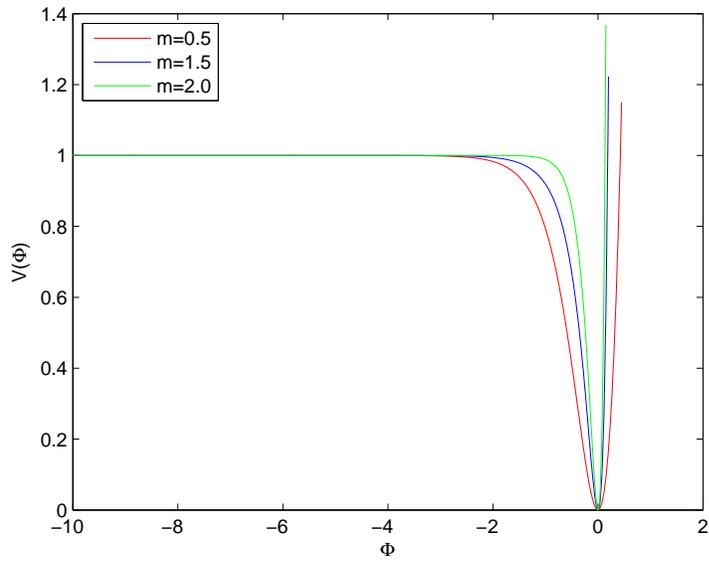}}
\caption{The potential of Eq. (\ref{pot2})
normalized to its value at $\Phi\longrightarrow -\infty$,
namely $V_{\infty}=V_{0}/2^{4n}$,
for $n=1$ and for various values of the parameter $m$.
\label{fig2}}
\end{figure}

In the present paper we consider the potential ($-\infty<\Phi\leq 0$)
\bea
\label{pot2}
V(\Phi)=V_{0}\frac{[tanh(\Phi/\sqrt{6\alpha})]^{2n}}{[1-tanh(\Phi/\sqrt{6\alpha})]^{2m}}
\eea
This is a generalization of the potential proposed in Fig. 2 of Ref. (\citen{lind1}),
where the parameter $\alpha$ is introduced, which is inversely proportional
to the curvature of the inflaton K\"{a}hler manifold \cite{kall1}. The parameters
$m,\; n$ are not necessarily integers. As in the previous case of the
Starobinsky model the slow-roll regime corresponds to the case of
$\Phi\longrightarrow -\infty$ where $F_{\hat{R}}=2e^{-\Phi/\sqrt{3}}\gg 1$.
The choice of the above potential for our study is partially based on the fact that
it is quite generic, possesses a large horizontal flat plateau for large negative
$\Phi-$values for the slow-roll inflation and possesses many limiting
cases as special cases, for example
the Starobinsky model \cite{star}, or the Higgs inflationary model \cite{capo1},
and so on. The potential of Eq. (\ref{pot2})
is shown in Figs. (\ref{fig1})-(\ref{fig2}).

Using the expansions (for $z:=e^{\sqrt{2/3\alpha}\Phi}\ll 1$)
\bea
\label{expa}
(1-z)^{N}\simeq 1-Nz+\frac{N(N-1)}{2}z^{2}\nn \\
\left(\frac{1}{1+z}\right)^{N}\simeq 1-Nz+\frac{N(N+1)}{2}z^{2}
\eea
we obtain in the slow-roll regime
\bea
\label{pot3}
V(\Phi)\simeq \frac{V_{0}}{2^{2m}}
\left[1-A(n,m)z+B(n,m)z^{2}\right]
\eea
where $A(n,m):=2(2n-m)$ and $B(n,m):=8n^{2}-8nm+2m^{2}-m$. From
Eq. (\ref{rel1}) we then obtain
\bea
\label{frg2}
\hat{R}F_{\hat{R}}-\frac{V_{0}}{2^{2m}}
\left[4\left(\frac{F_{\hat{R}}}{2}\right)^{2}+2A\left(\sqrt{\frac{2}{\alpha}}-2\right)\left(\frac{F_{\hat{R}}}{2}\right)^{\left(2-\sqrt{\frac{2}{\alpha}}\right)}\right.\nn
\\
\left.-4B\left(\sqrt{\frac{2}{\alpha}}-1\right)\left(\frac{F_{\hat{R}}}{2}\right)^{\left(2-2\sqrt{\frac{2}{\alpha}}\right)}\right] =0
\eea
This equation will be needed in the analysis that follows in section 5.

In the main part of the analysis that will follow we will be concerned
with the large curvature limit $(\alpha\longrightarrow 0)$. Also,
although the analysis that follows can be generalized to arbitrary values
of $m,\; n$, from now on we will focus on the specific subclass of models
where ${\bf m=2n}$. This is an exceptional and interesting case that is trackable
for analytical treatment. Then $A=0$, $B=-2n$ and we have
from Eq. (\ref{frg2})
\bea
\label{frg3}
\hat{R}-c_{0}F_{\hat{R}}-c_{1}F_{\hat{R}}^{-\delta}=0
\eea
where $c_{0}:=\frac{V_{0}}{2^{4n}}$, $c_{1}:=c_{0}2^{(\delta+1)}n^{(\delta -1)}$
and $\delta:=2\sqrt{\frac{2}{\alpha}}-1\geq 0$.

\section{Cosmological Parameters in the Einstein frame}

For the potential of Eq. (\ref{pot2}) and for $m=2n$ one can calculate the
cosmological observables of the spectral index of primordial curvature
perturbations $n_{s}$ and the scalar-to-tensor ratio $r$, as they occur in the
Einstein frame of Eq. (\ref{act3}), from
\bea
\label{obse}
n_{s}^{(E)}&=&1-6\epsilon+2\eta\nn \\
r^{(E)}&=&16\epsilon
\eea
where the slow-roll parameters are given by
\bea
\label{slow}
\epsilon&:=&\frac{1}{2}\left(\frac{V^{'}}{V}\right)^{2}\nn \\
\eta&:=&\frac{V^{''}}{V}
\eea
The calculation is exact as it was done in Eqs. (5.2)-(5.4)
of Ref. \citen{kall1}, {\it without} invoking the slow-roll hypothesis,
but only assuming that inflation ends when $\epsilon\simeq 1$. The
results are as follows: We define the number of e-foldings $N$ as
\bea
\label{efol}
N:=\int_{\Phi_{i}}^{\Phi_{e}}\frac{V(\Phi)}{V^{'}(\Phi)}d\Phi
\eea
where $\Phi_{i},\; \Phi_{e}$ are the initial and end values of the
inflaton scalar field. Along with the definitions $\xi:=tanh(\Phi/\sqrt{6\alpha})$ and
$g:=\frac{\sqrt{3\alpha}}{2n}$ we obtain
\bea
\label{efol1}
N=\frac{3\alpha}{4n}
\left[\frac{\Phi_{i}-\Phi_{e}}{\sqrt{6\alpha}}+\frac{1}{1+\xi_{e}}-\frac{1}{1+\xi_{i}}-\frac{1}{(1+\xi_{e})^{2}}+\frac{1}{(1+\xi_{i})^{2}}\right]
\eea
On the other hand from the requirement that inflation ends when $\epsilon\simeq 1$,
we obtain from the first of Eqs. (\ref{slow})
\bea
\label{efol2}
\xi_{e}=[-(1+g)+\sqrt{g(g+2)}]<0
\eea
Now remembering that $y=tanh^{-1}(x)=\frac{1}{2}ln|\frac{1+x}{1-x}|$ and
observing that in the limit of $\xi_{i}\longrightarrow -1+0$ the sixth term
in Eq. (\ref{efol1}) dominates the with respect to the fourth and the second terms,
we obtain
\bea
\label{efol3}
N=\frac{3\alpha}{4n}
\left[-\frac{1}{2}ln\left(\frac{1+\xi_{e}}{1-\xi_{e}}\right)+\frac{1}{(1+\xi_{i})^{2}}+\frac{1}{1+\xi_{e}}-\frac{1}{(1+\xi_{e})^{2}}\right]
\eea
Now using Eq. (\ref{efol2}) into Eq. (\ref{efol3}) and defining
\bea
\label{defs}
f(N,n,\alpha )&:=&\frac{4nN}{3\alpha}-\frac{1}{4}\left[ln\left(1+\frac{2}{g}\right)-\frac{2}{g}\right]\nn
\\
\xi_{o}&:=&\frac{1}{\sqrt{f}}-1
\eea
we obtain after a lengthy and careful calculation that
\bea
\label{inds}
\epsilon&=&\frac{3n^{2}}{\alpha}\frac{(1+\xi_{o})^{4}}{\xi_{o}^{2}}\nn
\\
\eta&=&\left[\frac{2n(2n-1)}{6\alpha}\frac{(1-\xi_{o}^{2})^{2}}{\xi_{o}^{2}}+\frac{16n^{2}}{6\alpha}\frac{(1+\xi_{o})^{2}}{\xi_{o}}\right]
\eea
and the cosmological observation parameters are given by Eq. (\ref{obse}),
using Eqs. (\ref{inds}).

\section{Cosmological Parameters in the Jordan frame}

It can be easily shown that an approximate
solution to Eq. (\ref{frg3}), valid in the slow-roll
regime $(F_{\hat{R}}\gg 1)$ is given by
\bea
\label{frg4}
F_{\hat{R}}&=&\frac{1}{c_{0}}\hat{R}-c_{1}c_{0}^{(\delta-1)}\hat{R}^{-\delta}\nn
\\
F(\hat{R})&=&\frac{\hat{R}^{2}}{2c_{0}}+\frac{c_{1}c_{0}^{(\delta -1)}}{(\delta-1)}\hat{R}^{(1-\delta)}+\Lambda
\eea
where $\Lambda$ is a positive cosmological constant. Now varying the action
of Eq. (\ref{act1}) we obtain\cite{capo}
($R$ is given by Eq. (\ref{ricci}) where $H=\dot{a}/a$ is the Hubble parameter and we
drop hats for simplicity, since it is clear that we work in the Jordan frame
of Eq. (\ref{act1}))
\bea
\label{feqs}
6H^{2}F_{R}&=&RF_{R}-F-6H\dot{R}F_{RR}\nn \\
-(2\dot{H}+3H^{2})F_{R}&=&F_{RRR}(\dot{R})^{2}+2HF_{RR}\dot{R}+F_{RR}\ddot{R}+\nn
\\
&+&\frac{F-RF_{R}}{2}
\eea
These reproduce Eq. (31) of Ref. (\citen{oik1}) in the proper limit.
In the slow-roll regime, where $(\dot{H}\ll H^{2})$ we may approximate
\bea
\label{appr}
R^{-\delta}=(12H^{2})^{-\delta}\left(1+\frac{\dot{H}}{2H^{2}}\right)^{-\delta}\simeq
\frac{1}{(12H^{2})^{\delta}}\left(1-\delta\frac{\dot{H}}{2H^{2}}\right)
\eea
Using Eqs. (\ref{frg4}) we obtain then from the first of Eqs. (\ref{feqs})
we obtain after a slightly lengthy calculation
\bea
\label{feq1}
36H\ddot{H}+\Lambda c_{0}-18\dot{H}^{2}+108H^{2}\dot{H}+\nn \\
+\frac{6c_{1}c_{0}^{\delta}}{(12H^{2})^{\delta}}
\left[\frac{(\delta+1)}{(\delta-1)}H^{2}+\frac{3}{2}\delta\dot{H}-\delta(\delta-1)\frac{\dot{H}^{2}}{H^{2}}\right]=0
\eea
This generalizes Eq. (32) of Ref. (\citen{oik1}). It is quite interesting and
unlike the case of Eq. (32) of Ref. (\citen{oik1}), where in order to obtain
a non-trivial solution they had to take the derivative of Eq. (31) in order to obtain
Eq. (36) and finally Eq. (37) (see Ref. (\citen{oik1})), that, in our case,
an {\it exact} solution to Eq. (\ref{feq1}) is given by
(in the slow-roll regime where $H_{1}\ll H_{0}$)
\bea
\label{solu}
H(t)&=&H_{0}-H_{1}(t-t_{k})
\eea
where $H_{0},\; t_{k}$ are arbitrary integration constants of Eq. (\ref{feq1}).
We assume that the cosmological constant is given by
$\Lambda c_{0}=108H_{0}^{2}H_{1}>0$. Then the third and the last
three terms of Eq. (\ref{feq1}) are equated to zero and this gives
\bea
\label{H1}
H_{1}&=&\frac{1}{2k_{0}}\left[-k_{1}+\sqrt{k_{1}^{2}-4k_{0}k_{2}}\right]>0\nn
\\
k_{0}&=&3\left[1+\frac{4\delta (\delta-1)c_{1}c_{0}^{\delta}}{(12H_{0}^{2})^{(\delta+1)}}\right]\nn
\\
k_{1}&=&\frac{3}{2}\frac{\delta c_{1}c_{0}^{\delta}}{(12H_{0}^{2})^{\delta}}\nn
\\
k_{2}&=&-\left(\frac{\delta +1}{\delta -1}\right)H_{0}^{2}\frac{c_{1}c_{0}^{\delta}}{(12H_{0}^{2})^{\delta}}
\nn \\
c_{0}&=&\frac{V_{0}}{2^{4n}}\nn \\
c_{1}&=&c_{0}n^{(\delta -1)}2^{(\delta +1)}
\eea

Specifically the time $t_{k}$ is assumed to be the time where the
horizon crossing for the comoving wavenumber $k=a(t)H(t)$ occurred.
For the action of Eq. (\ref{act1}) the cosmological observables corresponding
to Eq. (\ref{obse}) are given by (see Refs. (\citen{oik1}) and (\citen{hnoh}))
\bea
\label{obse1}
n_{s}&\simeq& 1-6\epsilon_{1}-2\epsilon_{4}\simeq 1-\frac{2\dot{\epsilon}_{1}}{H(t)\epsilon_{1}}\nn
\\
r&=&48\epsilon_{1}^{2}
\eea
where
\bea
\label{obse2}
\epsilon_{1}=-\frac{\dot{H}}{H^{2}},\; \; \epsilon_{2}=0,\; \;
\epsilon_{3}\simeq \epsilon_{1},\; \; \epsilon_{4}\simeq -3\epsilon_{1}+\frac{\dot{\epsilon}_{1}}{H(t)\epsilon_{1}}
\eea

If we assume that the slow-roll regime (and essentially the inflation also)
ends when $\epsilon_{1}\simeq {\cal O}(1)$, at $t=t_{f}$, so that $H(t_{f}):=H_{f}$
we have $H_{f}=\sqrt{H_{1}}$ and also \\
$(t_{f}-t_{k})\simeq (H_{0}/H_{1})$. Defining the number of e-foldings of inflation
as
\bea
\label{efol4}
N=\int_{t_{k}}^{t_{f}}H(t)dt
\eea
we end up with
\bea
\label{tftk}
(t_{f}-t_{k})\simeq \frac{2N}{H_{0}}
\eea
Thus finally the cosmological observables of Eq. (\ref{obse1}), as they occur in the
Jordan frame, are given by
\bea
\label{obse3}
n_{s}^{(J)}&\simeq &1-\frac{4H_{1}}{\left(H_{0}-\frac{2H_{1}N}{H_{0}}\right)^{2}}\nn
\\
r^{(J)}&\simeq&\frac{48H_{1}^{2}}{\left(H_{0}-\frac{2H_{1}N}{H_{0}}\right)^{4}}
\eea
Although these equations are similar in form with Eq. (50) of Ref. (\citen{oik1}),
they are essentially different in the fact that they depend additionally on the
parameter $\alpha$ and on the freely specified parameter $H_{0}$. Hereafter we choose
$H_{0}=1.$ In all the numerical examples below we found that in Eq. (\ref{H1}),
$H_{1}\leq 0.02$ in practically all the cases, so that the slow-roll approximation
($H_{1}\ll H_{0}$) is indeed satisfied in all of the relevant cases.

\section{Discussion}

\begin{figure}[h!]
\centerline{\includegraphics[width=10.8cm]{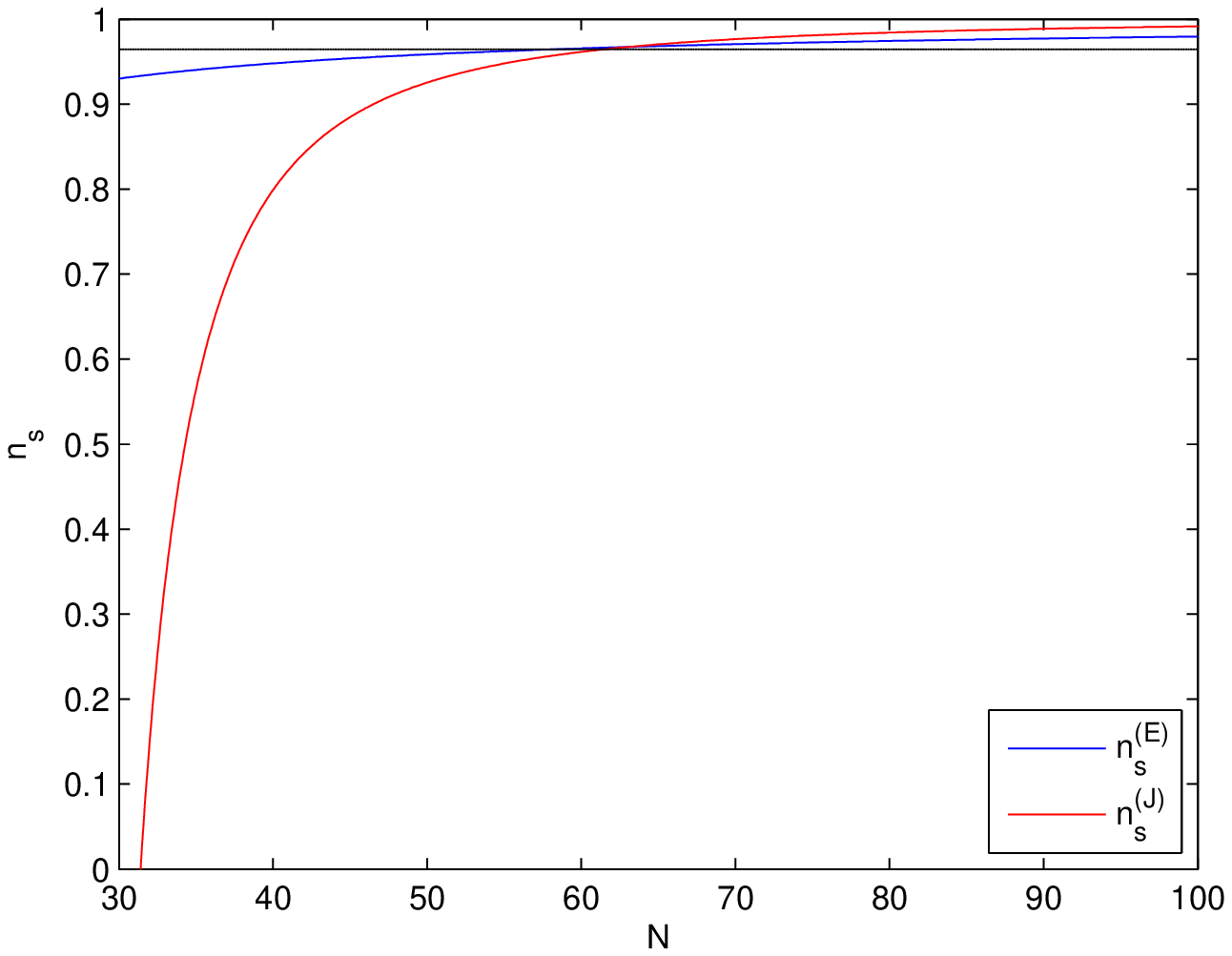}}
\caption{The spectral index of primordial curvature perturbations $n_{s}$,
in the Einstein and Jordan frames, as a function of the number of e-foldings $N$,
namely that of Eqs. (\ref{obse}) and (\ref{obse3}),
for $n=1,\; m=2n$, $\alpha=0.0625$. In black it is shown the constant value
of Eq. (\ref{ns}).
\label{fig3}}
\end{figure}

\begin{figure}[h!]
\centerline{\includegraphics[width=10.8cm]{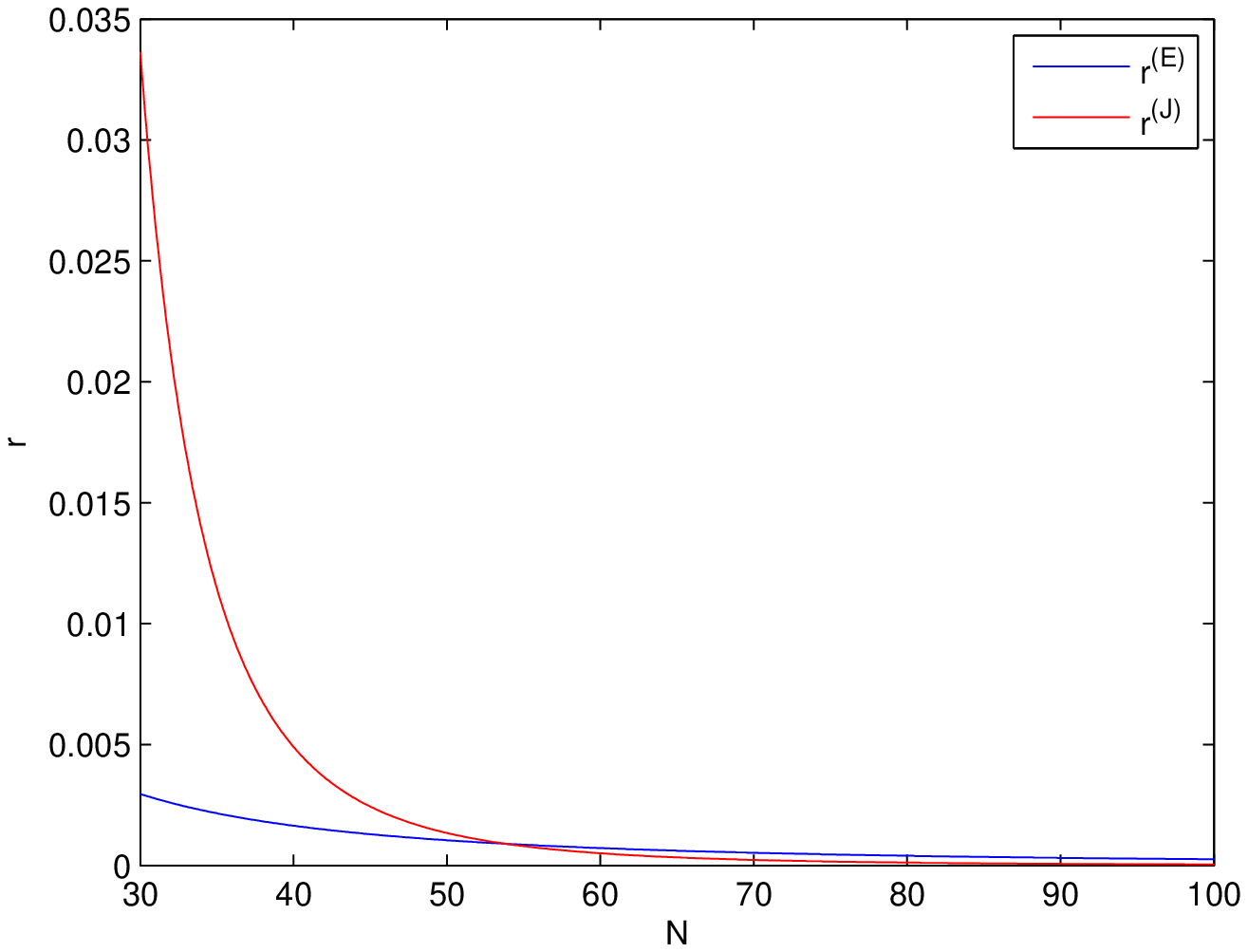}}
\caption{The tensor-to-scalar ratio $r$,
in the Einstein and Jordan frames, as a function of the number of e-foldings $N$,
namely that of Eqs. (\ref{obse}) and (\ref{obse3}),
for $n=1,\; m=2n$, $\alpha=0.09$.
\label{fig4}}
\end{figure}

\begin{figure}[h!]
\centerline{\includegraphics[width=10.8cm]{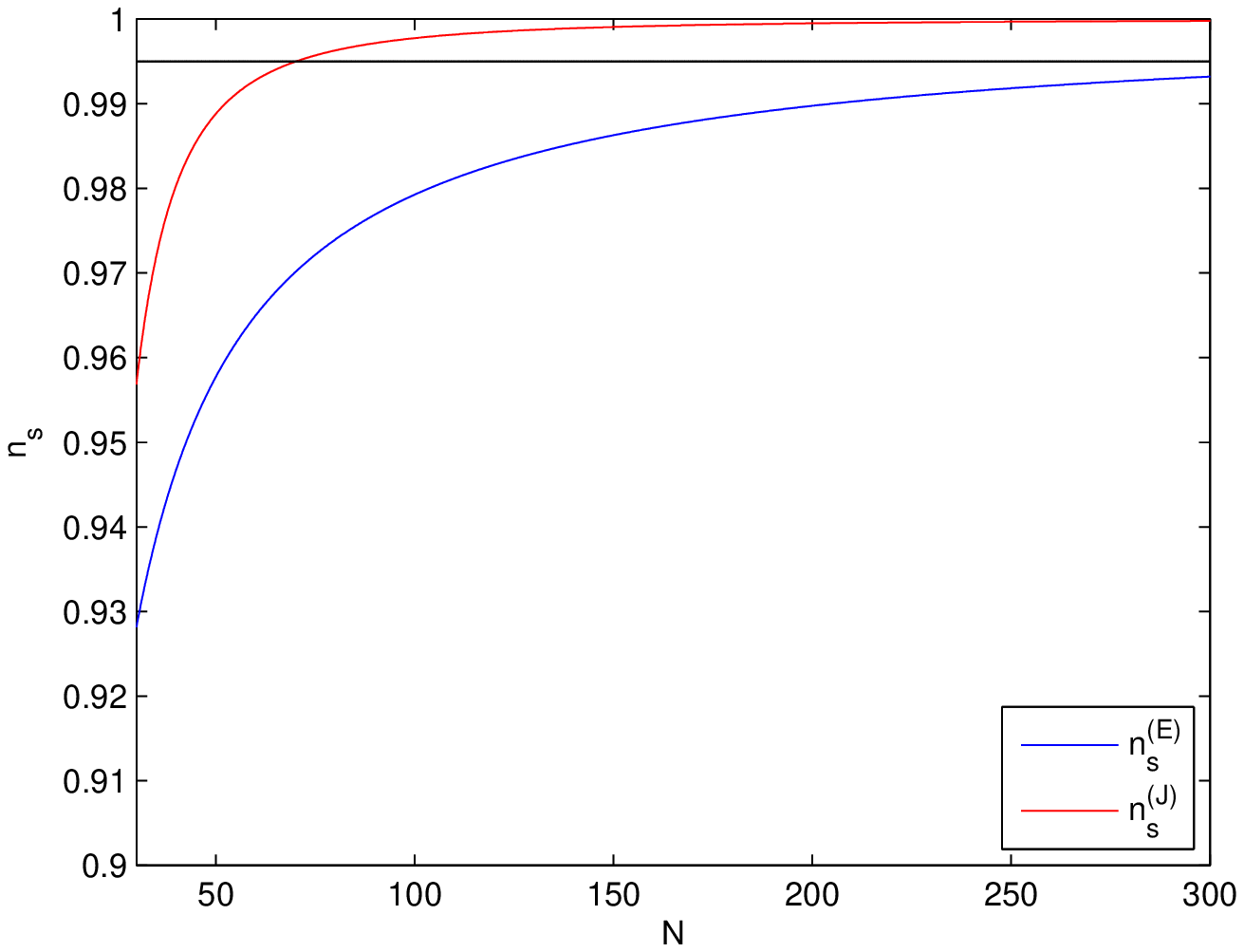}}
\caption{The spectral index of primordial curvature perturbations $n_{s}$,
in the Einstein and Jordan frames, as a function of the number of e-foldings $N$,
namely that of Eqs. (\ref{obse}) and (\ref{obse3}),
for $n=1,\; m=2n$, $\alpha=0.125$. In black it is shown the constant value
of $n_{sa}=0.995$.
\label{fig3a}}
\end{figure}

\begin{figure}[h!]
\centerline{\includegraphics[width=10.8cm]{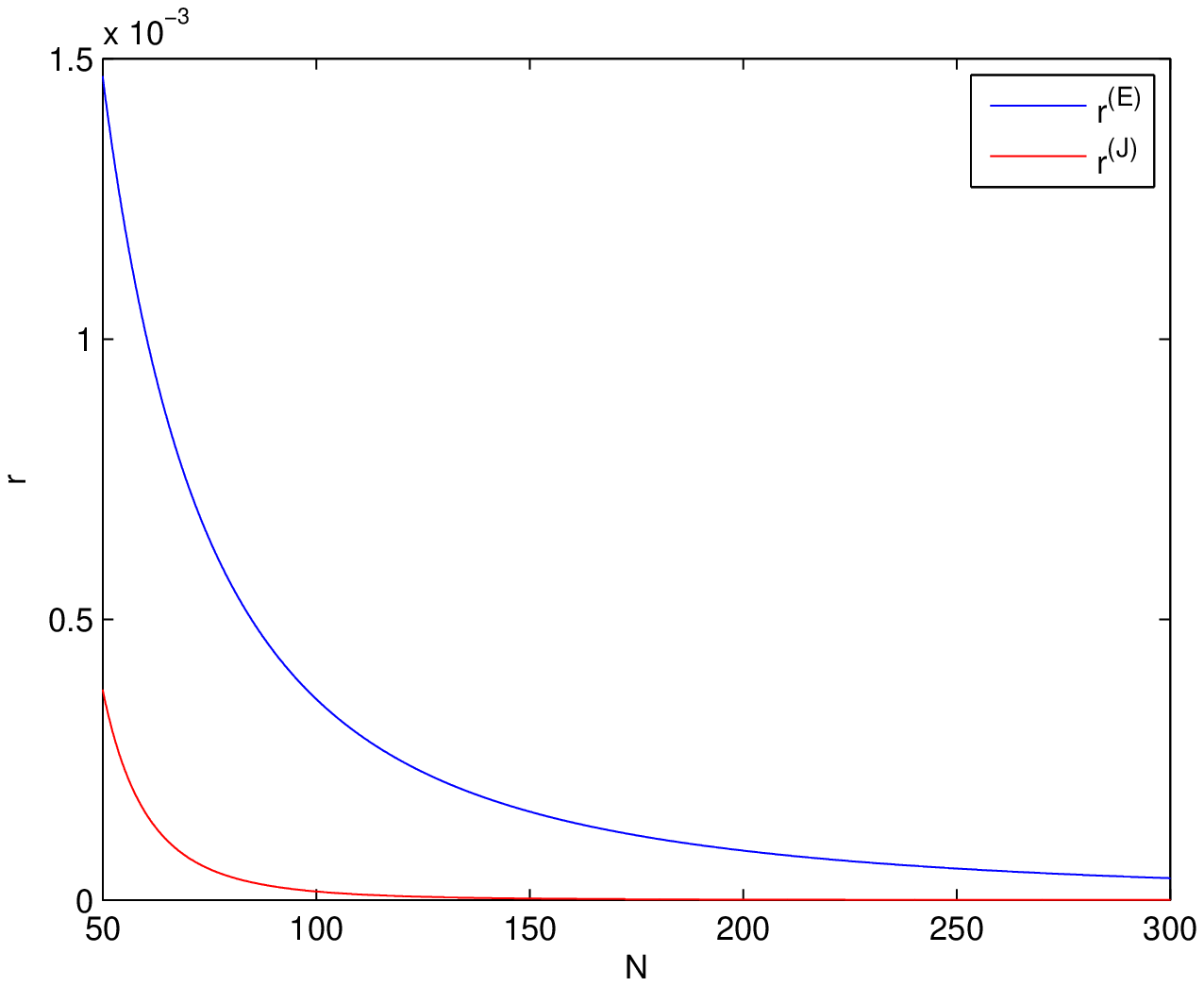}}
\caption{The tensor-to-scalar ratio $r$,
in the Einstein and Jordan frames, as a function of the number of e-foldings $N$,
namely that of Eqs. (\ref{obse}) and (\ref{obse3}),
for $n=1,\; m=2n$, $\alpha=0.125$.
\label{fig4a}}
\end{figure}

\begin{figure}[h!]
\centerline{\includegraphics[width=10.8cm]{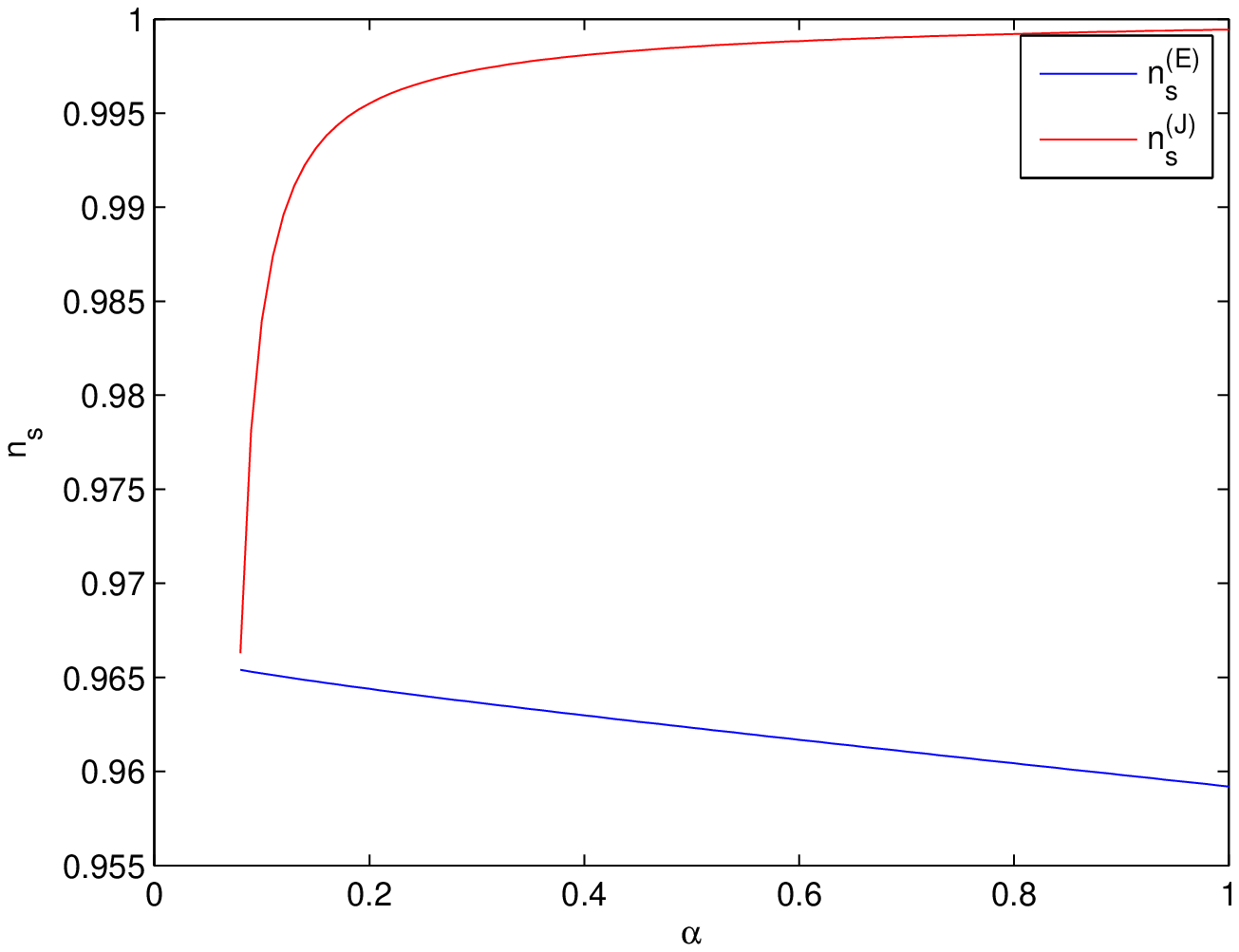}}
\caption{The spectral index of primordial curvature perturbations $n_{s}$,
in the Einstein and Jordan frames, as a function of the parameter $\alpha$,
namely that of Eqs. (\ref{obse}) and (\ref{obse3}),
for $N=60$ and $n=1,\; m=2n$.
\label{fig5}}
\end{figure}

\begin{figure}[h!]
\centerline{\includegraphics[width=10.8cm]{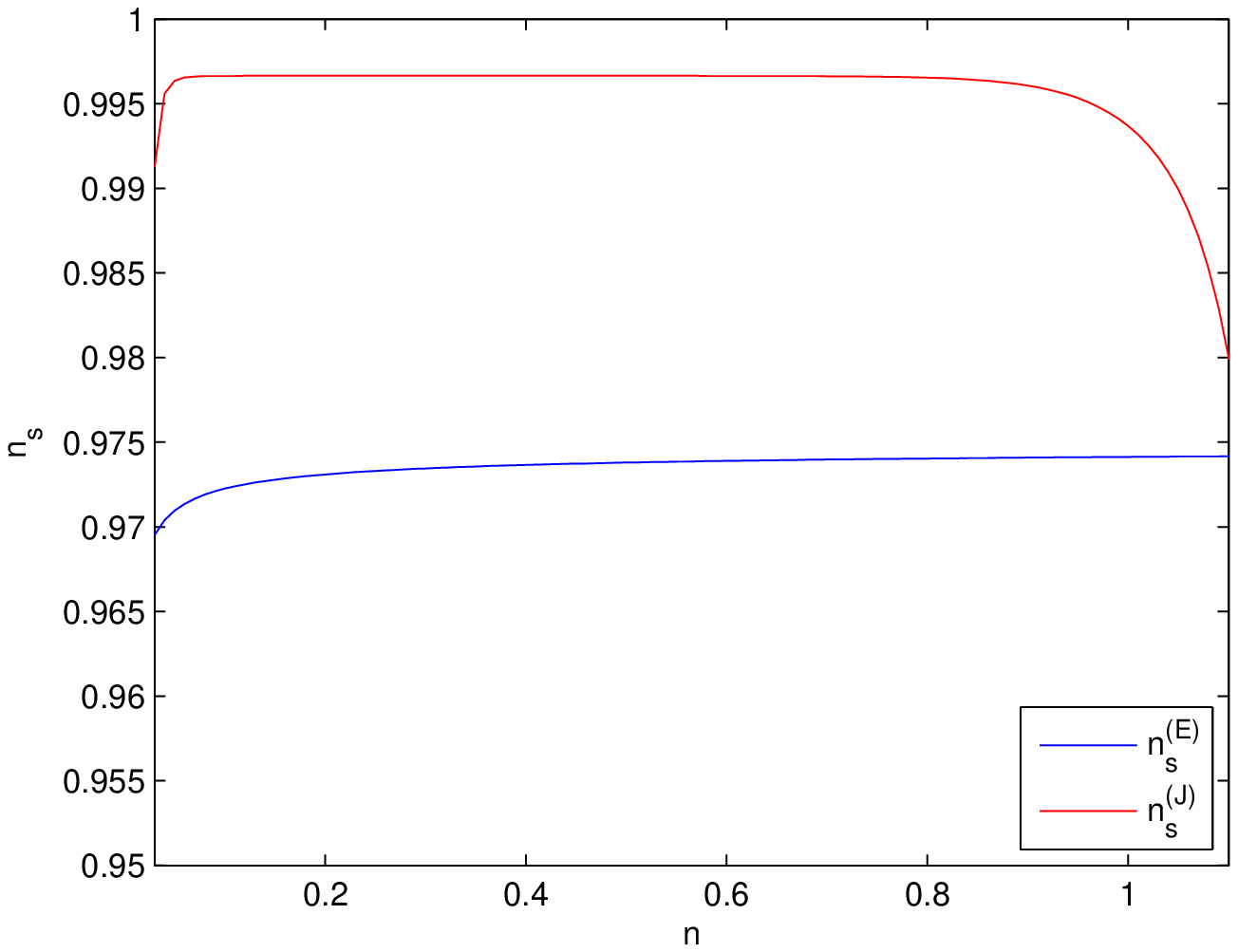}}
\caption{The spectral index of primordial curvature perturbations $n_{s}$,
in the Einstein and Jordan frames, as a function of the parameter $n$,
namely that of Eqs. (\ref{obse}) and (\ref{obse3}),
for $m=2n$, $N=80$ and $\alpha=0.09$.
\label{fig6}}
\end{figure}

In this section we are ready to compare the observational indices as
they occur in the Einstein and Jordan frames of Eqs. (\ref{obse}) and
Eqs. (\ref{obse3}) respectively. In order to make compatible our
results with Eqs. (15), (22) of Ref. \citen{ade},
namely
\bea
\label{ns}
n_{s}&=&0.9645\pm 0.0049\nn \\
r&<&0.10
\eea
we choose to have $n_{s}^{(E)}(N=60)=0.9645$ in Fig. \ref{fig3}.
This was easily achieved by choosing $\alpha=0.0625$, as it is referred to in
the caption of Fig (\ref{fig3}). This is shown by the black horizontal line.
Also we observe that the two curves are practically
asymptote to this value for all values of $N\geq 50$ (to within errors due to the
use of the slow-roll approximation), namely the two sets of observables
coincide. This is the main result of the present paper that generalizes the results
of Ref. (\citen{oik1}).
The curves of Fig. \ref{fig3}, as has been referred to,
correspond to the case of $\alpha=0.0625,\; \; n=1$.
This is in our viewpoint a novel result, unlike the case of Ref. (\citen{oik1}),
where the observational indices in the two frames coincide only in the
large-$N$ limit and in the small$-\alpha$ limit. In our case this happens also
for $\alpha\leq 0.1$ also for moderate $N-$values.
In the same manner we obtain also Fig. \ref{fig4},
where we find that $r^{(E)}(N=60)\simeq 0.00072038<0.10$,
for the parameter value of $V_{0}=45$. When the value of the $\alpha-$parameter
increases however, (namely for $\alpha=0.125$) we obtain the curves of
Figs. (\ref{fig3a})-(\ref{fig4a}). Here the observational indices
in the two frames do not asymptote to a common value unless the number of
e-foldings is larger than $N\geq 300$. Therefore the two sets of observational indices,
corresponding to the two frames coincide for practically all relevant values
of the number of e-foldings $N$, when $\alpha\leq 0.1$ and $n\simeq 1$.
This is also evident from Figs. (\ref{fig5})-(\ref{fig6}). It is quite possible,
although difficult to ascertain analytically, that without invoking the
condition $m=2n$ in the potential of Eq. (\ref{pot2}), equivalence of the
two frame descriptions would occur for an even vaster range of parameter values.

Regarding the crucial issue of whether these attractors and the
observational indices connected with them can be used to
distinguish between the two frames (namely the Einstein and Jordan frames)
the author considers this to be a very deep question and a definite
answer cannot so easily be given. However according to the authors viewpoint
and relevant work on this subject (see Refs. \citen{capo2}, \citen{capo1}, \citen{capo3}),
in the case considered in this paper, although the two frames are
mathematically equivalent, as they are connected by a conformal transformation,
there exists a physical non-equivalence of the two frames, and by using the
results obtained here, regarding the observational indices, one may be
able to distinguish between the two frames.

Since it is quite difficult to obtain a definite answer\cite{oik1},
for the most generic case, regarding the equivalence of the
descriptions of the observational indices in the two
frames (Einstein and Jordan frames), it would be interesting to try
to check this postulate for more realistic and general inflationary potentials,
as those for example suggested by appropriate limits of certain
supergravity and/or string theory actions (see Ref. (\citen{lind2})
and references therein). Work along these lines is in progress.


\section*{Acknowledgments}

The author would like to acknowledge useful discussions with
Dr. K. Kleidis and other colleagues in the Technological Education
Institute (T.E.I.) of Central Macedonia, Greece.  The author would like
to acknowledge the useful criticisms and remarks of the anonymous
referees that helped improve the paper.





\end{document}